\documentclass[amsmath,amssymb,twocolumn,preprintnumbers,nofootinbib,prd]{revtex4-1}
\usepackage{amsmath,amssymb,graphicx,booktabs,bm,psfrag,slashed}

\usepackage[usenames,dvipsnames]{color}
\definecolor{darkblue}{rgb}{0,0,0.5}
\definecolor{darkgreen}{rgb}{0.1,0,0.3}
\definecolor{darkred}{rgb}{0.6,0,0}
\usepackage[bookmarks=true,bookmarksnumbered=true,colorlinks=true,urlcolor=darkblue,citecolor=darkgreen,linkcolor=darkred,breaklinks=true]{hyperref} 
\usepackage{bbm}
\usepackage{xspace}
\usepackage{mhchem}
\usepackage{mathtools}
\usepackage[normalem]{ulem}
\def\bs{\ensuremath\boldsymbol}

\newcommand{\Ubl}{$U(1)_{B-L}$\xspace}
\newcommand{\Ubli}[1]{$U(1)_{B-3L_\text{#1}}$\xspace}
\newcommand{\coherent}{CE$\nu$NS\xspace}

\begin{document}

\title{On the Flavour Structure
 of Anomaly-free Hidden Photon Models}

\preprint{IPPP/20/59}

\author{Martin Bauer$^1$}
\author{Patrick Foldenauer$^1$}
\author{Martin Mosny$^2$}

\affiliation{$^1$Institute for Particle Physics Phenomenology, Department of Physics, Durham University, Durham, DH1 3LE, United Kingdom}
\affiliation{$^2$DAMTP, University of Cambridge, Wilberforce Road, Cambridge, CB3 0WA, United Kingdom}

\begin{abstract}
Extensions of the Standard Model with an Abelian gauge group are constrained by gauge anomaly cancellation, so that only a limited number of possible charge assignments is allowed without the introduction of new chiral fermions. For flavour universal charges, couplings of the associated hidden photon to Standard Model fermions are flavour conserving at tree-level. We show explicitly that even the flavour specific charge assignments allowed by anomaly cancellation condition lead to flavour conserving tree-level couplings of the hidden photon to quarks and charged leptons \emph{if} the CKM or PMNS matrix can be successfully reconstructed. Further, loop-induced flavour changing couplings are strongly suppressed. As a consequence, the structure of the Majorana mass matrix is constrained and flavour changing tree-level couplings of the hidden photon to neutrino mass eigenstates are identified as a means to distinguish the $U(1)_{B-L}$  gauge boson from any other anomaly-free extension of the Standard Model without new chiral fermions. We present a comprehensive analysis of constraints and projections for future searches for a $U(1)_{B-3L_i}$ gauge boson, calculate the reach of resonance searches in $B$ meson decays and comment on the implications for non-standard neutrino interactions. 
\end{abstract}

\maketitle

\section{ Introduction} 
Light new physics can arise in many extensions of the Standard Model of particle physics (SM). Depending on mass and couplings, new light vector bosons can be constrained by low-mass resonance searches~\cite{Athanassopoulos:1997er, Aguilar:2001ty, Batley:2015lha,Aubert:2009cp, Essig:2009nc, Merkel:2011ze, Merkel:2014avp,Archilli:2011zc, Babusci:2012cr, Babusci:2014sta, Curtin:2014cca, Lees:2014xha, Ilten:2015hya, Ilten:2016tkc, Anastasi:2016ktq, Inguglia:2016acz, Aaij:2017rft, Evans:2017lvd, Gligorov:2017nwh},
 missing energy signatures~\cite{Kahn:2012br, Gninenko:2014pea, Kaneta:2016uyt, Araki:2017wyg, Lees:2017lec, Gninenko:2018tlp}, 
 scattering experiments~\cite{Riordan:1987aw, Bjorken:1988as, Bross:1989mp, Davier:1989wz, MeijerDrees:1992kd, Bjorken:2009mm, Essig:2010gu, Blumlein:2011mv, Abrahamyan:2011gv, Andreas:2012mt, Blumlein:2013cua, Altmannshofer:2014pba, Alekhin:2015byh, Bilmis:2015lja, Kamada:2015era, Araki:2015mya, DiFranzo:2015qea, Banerjee:2016tad, Ge:2017poy} and flavour observables~\cite{Bertl:1985mw, Alam:1995mt, Blondel:2013ia, Echenard:2014lma, Altmannshofer:2014cfa, Wise:2018rnb}. 
 Because of the SM flavour structure, flavour observables provide a particularly powerful probe. There is a priori no reason to assume that any extension of the SM respects the SM flavour symmetry in the absence of a mechanism that gives rise to minimal flavour violation. Such a mechanism is for example realised for new gauge bosons that only couple to the SM through kinetic mixing with the hypercharge gauge boson~\cite{Holdom:1985ag, He:1990pn, He:1991qd, Pospelov:2008zw}. We discuss in how far this mechanism is realised for new gauge groups with charged SM matter. The allowed charge assignments in such new $U(1)$ extension are strongly constrained by the requirement of anomaly cancellation and the structure of the CKM and PMNS mixing matrices. Together, these constraints leave a limited number of global symmetries of the SM that can be gauged with the addition of only 3 right-handed neutrinos: $U(1)_{B-L},  U(1)_{B-3L_i}$, and combinations of these groups. We will refer to the corresponding gauge groups as \emph{minimal} anomaly-free $U(1)$ extensions of the SM and classify the possible Majorana matrix textures that are allowed by the masses and mixing angles of the SM fermions. We demonstrate that loop-induced flavour changing couplings of the new gauge bosons to quarks and charged leptons are protected by a GIM mechanism even for flavour non-universal gauge couplings and estimate the reach of dedicated resonance searches for new gauge bosons in flavour changing meson decays. In contrast, flavour changing couplings to neutrino mass eigenstates are induced at tree-level for all gauge groups with non-universal neutrino charges. We discuss the implications of such non-universal neutrino charges for non-standard neutrino interactions and present a comprehensive analysis of current and future experiments searching for a new gauge boson in the case of the three $U(1)_{B-3L_i}$ groups.

\section{The Flavour Structure of Hidden Photon Couplings}

\begin{table*}[t!]
\centering\renewcommand\arraystretch{1.3}
\begin{tabular}{p{2.3cm} c c}
\hline
\hline
 Anomaly & Charge combinations  & With Yukawa constraints\\
\hline
 $U(1)^3_X$ & $2X_L^3+ 6 X_Q^3  - X_\ell^3 - X_\nu^3 - 3(X_u^3 + X_d^3) $& $X_L^3-X_\nu^{3}$ \\
 $U(1)^2_X U(1)_Y$ & $2 Y_L X_L ^2+ 6Y_Q X_Q^2 - Y_\ell X_\ell^2 - Y_\nu X_\nu^2 - 3( Y_u X_u^2 +  Y_d X_d^2) $& $0$ \\
 $ U(1)_XU(1)_Y^2$ & $2 Y_L^2 X_L + 6 Y_Q^2 X_Q - Y_\ell^2 X_\ell -  Y_\nu^2 X_\nu - 3( Y_u^2X_{u} + Y_d^2X_{d}) $ &$-\frac{1}{2}\left(X_L+3X_Q\right)$\\
 $ SU(3)^2U(1)_X$ & $ 2X_Q - X_{u} - X_{d}  $&$0$ \\
 $ SU(2)^2U(1)_X$ & $2X_L  + 6X_Q  $ &$2X_L  + 6X_Q  $\\
 $ \text{grav}^2U(1)_X$ & $2X_L + 6X_Q - X_\ell - X_\nu - 3(X_u + X_d) $&$X_L-X_\nu$ \\
\end{tabular}
\vspace{.2cm}
\caption{Constraints on the $U(1)_X$ and $U(1)_Y$ charges with the hypercharge denoted by $Y_\psi$ and the $U(1)_X$ charges defined in \eqref{eq:Xcharge}.}
\label{anomalyc1}
\end{table*}

We consider the SM extended by a new $U(1)_X$ gauge group and three neutrinos transforming as singlets under $SU(3)_C\times SU(2)_L\times U(1)_Y$ and no additional fermions charged under $U(1)_X$.  Three sets of constraints on the possible $U(1)_X$ charges of the SM fermions arise from anomaly cancellation, from the observed quark and lepton masses and from the requirement to reproduce the structures of the CKM and PMNS matrices. The couplings of the $X$ gauge boson to SM fermions are defined by 
\begin{align}
\mathcal{L}_X\ni& X_\mu \sum_\psi \bar \psi \,i g_X Q_\psi\gamma^\mu \psi\,,
\end{align}
where the fermion sum extends over $SU(2)_L$ doublets $\psi=Q,L$ and singlets $\psi=u,d,\ell,\nu$, which each denote vectors in flavour space. In the interaction basis the $U(1)_X$ charge matrices are flavour diagonal and we define $Q_\psi = \text{diag}\,(q_{\psi_1}, q_{\psi_2}, q_{\psi_3}) \equiv T_\psi$.  Sums over fermion charges can then be compactly expressed by writing
\begin{align}\label{eq:Xcharge}
X_\psi^n &= \sum_{i}^3 (q_{\psi_i})^n\,,
\end{align} 
and $X_\psi\equiv X^1_\psi$.
Anomaly cancellation implies six conditions on the $U(1)_X$ and $U(1)_Y$ charges shown in Table~\ref{anomalyc1}. Together they allow for two solutions
\begin{align}\label{eq:Ancases}
&\text{I}:\, X_L=-X_\ell=\frac{1}{3}X_\nu=-\frac{3}{7}X_d=-\frac{5}{3}X_u=-3X_Q\,,\notag\\
&\text{II}:\, X_L=X_\ell=X_\nu=-3X_d=-3X_u=-3X_Q\,.
\end{align}

An independent set of constraints arises from the structure of the Yukawa couplings.
The three observed independent masses for the up-type and down-type quarks as well as for charged leptons constrain the structure on the Yukawa couplings defined by
\begin{align}\label{eq:Yukawa}
\mathcal{L}_Y=\frac{v}{\sqrt2}\sum_\psi \bar\psi y_\psi \psi \,.
\end{align}
If all fermion masses are generated by the Yukawa terms, all Yukawa matrices need to be rank-3. This implies that the $U(1)_X$ charges have to be permutations of each other $T_Q=\text{perm}(T_u)=\text{perm}(T_d)$ so that $X_u^n=X_d^n=X_Q^n\equiv X_\text{quarks}^n$. Similarly for charged leptons $T_L=\text{perm}(T_\ell)$ and $X_\ell^n=X_L^n\equiv X_\text{leptons}^n$. If there are three massive Dirac neutrinos the Yukawa structure also demands $X_\nu^n=X_L^n$, but if one neutrino is massless $Y_\nu$ can be rank-2 and this equation only holds due to the anomaly conditions. 

\subsection{Dirac neutrinos}

For three massive Dirac neutrinos, the combination of the constraints from anomaly cancellation and Yukawa matrices can then be expressed by the single equation
\begin{align}\label{eq:anomalycondition}
X_\text{leptons}+3X_\text{quarks}=0\,.
\end{align}
This is fulfilled by case II in \eqref{eq:Ancases}, but rules out case I. 

The Yukawa Lagrangian \eqref{eq:Yukawa} is structurally invariant under permutations of the charge tuples $T_\psi$ and we can set $T_Q = T_u = T_d$ and $T_L = T_e = T_\nu$ without loss of generality and we only need to consider three different tuples of charges: $(a,a,a)$, $(a,a,b)$ and $(a,b,c)$, referred to as the $a^3$, $a^2b$ or $abc$ classes, respectively.

Importantly, for all three classes flavour changing neutral currents are absent at tree-level. Upon rotating to the mass basis $y^\text{diag}_\psi = U_\psi y_\psi W_\psi^\dagger$ the couplings of the $U(1)_X$ gauge boson read $Q_\psi \to U_\psi Q_\psi U_\psi^\dagger$ for $\psi=Q,L$ and $Q_\psi \to W_\psi Q_\psi W_\psi^\dagger$ for $\psi=u,d,\ell,\nu$ and in all cases
\begin{align}\label{eq:com}
[Q_\psi, U_\psi]=[Q_\psi, W_\psi]=0\,.
\end{align}
For the $a^3$ class $Q_\psi\propto \mathbbm{1}$ and commutes with all matrices $U_\psi, W_\psi$. For the $a^2b$ class, the Yukawa matrices and the rotation matrices $U_\psi, W_\psi$ are block-diagonal and commute with $Q_\psi$, because each blocks acts on the submatrices of $Q_\psi$ that are proportional to the unit matrix. For the $abc$ class, the Yukawa matrices are already diagonal and $U_\psi, W_\psi\propto \mathbbm{1}$.

Finally, the three angles and the phase of the CKM and PMNS matrices need to be reproduced. The CKM matrix is defined by $V_\text{CKM}=U_u U_d^\dagger$ and the PMNS matrix for Dirac neutrinos is defined by $V_\text{PMNS}=U_\ell U_\nu^\dagger$. For the $a^3$ class $U_\psi$ are general unitary matrices, for $a^2b$ they are unitary block diagonal matrices, while for the $abc$ class they are complex diagonal matrices.  The anomaly condition \eqref{eq:anomalycondition} implies that the quark sector can only be charged under $U(1)_X$ if the lepton sector is charged as well. If all neutrinos are massive, the PMNS matrix excludes any class apart from the $a^3$ class. The anomaly condition further implies a relative charge between the lepton and quark sectors of $X_L = -\frac{1}{3}X_Q$, only allowing for the $U(1)_{B-L}$ gauge group. In the absence of additional contributions to the fermion mass terms such as Majorana mass terms or higher order operators $U(1)_{B-L}$ therefore is the only $U(1)$ extension allowed by all constraints. If instead one neutrino is massless, the PMNS matrix can never have the necessary number of degrees of freedom and any gauge group $U(1)_X$ with charged SM fermions and no additional chiral fermions is excluded. 
The presence of additional contributions to the mass terms will in general introduce tree-level flavour changing neutral current (FCNCs). In the following we classify the different coupling structures allowed in the presence of Majorana mass terms. \\

\subsection{Majorana neutrinos}
Majorana mass terms for the three right-handed neutrinos are of the form 
\begin{equation} \label{lag}
\mathcal L \ni 
- i  M_{ij}\nu_R^{Ti}\sigma_2 \nu_R^j - i C_{ij}^k\phi_k \nu_R^{Ti}\sigma_2 \nu_R^j + h.c.,
\end{equation}
where $M$ and $C^k$ are complex, symmetric matrices and $\phi_k$ are complex scalars, neutral under the SM gauge group, but charged under $U(1)_X$.  A vacuum expectation value for any of the scalars $\langle\phi_k\rangle$ contributes to the mass of the $U(1)_X$ gauge boson $X_\mu$ and generates contributions to the Majorana mass matrix that are otherwise forbidden by gauge invariance. The complete Majorana mass matrix is then given by
\begin{align}
M_{ij}^M = 2(M_{ij} +C^k_{ij}\langle \phi_k \rangle)\,.
\end{align}
Note that the textures of the matrices $M$ and $C^k$ are orthogonal in the sense that if an element $M_{ij}\neq 0$ the corresponding element $C^k_{ij}=0$ and vice versa.
In the seesaw limit, the mass of the active neutrinos are obtained by diagonalizing the matrix
\begin{equation} \label{seesawm}
m_\nu = -m^T (M^M)^{-1}m,
\end{equation}
where $m_{ij} = \frac{v}{\sqrt 2}y^\nu_{ij}$ are the Dirac mass terms.

For the seesaw limit (\ref{seesawm}) to work, the Majorana matrix must be rank-3 to be invertible which limits the allowed charge assignments. 
\begin{table}[t]
\begin{gather*}\addtolength{\jot}{1cm}
\bs A_1= \begin{pmatrix} 0 & 0 & \cdot \\0 & \cdot & \cdot \\ \cdot & \cdot & \cdot \end{pmatrix}, \ \ \ \ \bs A_2= \begin{pmatrix} 0 & \cdot & 0 \\ \cdot & \cdot & \cdot \\ 0 & \cdot & \cdot \end{pmatrix}; \\
\bs B_1= \begin{pmatrix} \cdot & \cdot & 0 \\ \cdot & 0 & \cdot \\ 0 & \cdot & \cdot \end{pmatrix}, \ \ \ \ \bs B_2= \begin{pmatrix} \cdot & 0 & \cdot \\0 & \cdot & \cdot \\ \cdot & \cdot & 0 \end{pmatrix}, \\ 
\bs B_3= \begin{pmatrix} \cdot & 0 & \cdot \\0 & 0 & \cdot \\ \cdot & \cdot & \cdot \end{pmatrix}, \ \ \ \ \bs B_4= \begin{pmatrix} \cdot & \cdot & 0 \\ \cdot & \cdot & \cdot \\ 0 & \cdot & 0 \end{pmatrix}; \\
\bs C= \begin{pmatrix} \cdot & \cdot & \cdot \\ \cdot & 0 & \cdot \\ \cdot & \cdot & 0 \end{pmatrix}; \ \ \ \
\bs D_1= \begin{pmatrix} \cdot & \cdot & \cdot \\ \cdot & 0 & 0 \\ \cdot & 0 & \cdot \end{pmatrix}, \ \ \ \ \bs D_2 =\begin{pmatrix} \cdot & \cdot & \cdot \\ \cdot & \cdot & 0 \\ \cdot & 0 & 0 \end{pmatrix}; \\
\bs E_1= \begin{pmatrix} 0 & \cdot & \cdot \\ \cdot & 0 & \cdot \\ \cdot & \cdot & \cdot \end{pmatrix}, \ \ \ \ \bs E_2= \begin{pmatrix} 0 & \cdot & \cdot \\ \cdot & \cdot & \cdot \\ \cdot & \cdot & 0 \end{pmatrix}, \ \ \ \ \bs E_3= \begin{pmatrix} 0 & \cdot & \cdot \\ \cdot & \cdot & 0 \\ \cdot & 0 & \cdot \end{pmatrix}; \\
\bs F_1= \begin{pmatrix} \cdot & 0 & 0 \\ 0 & \cdot & \cdot \\ 0 & \cdot & \cdot \end{pmatrix}, \ \ \ \ \bs F_2=\begin{pmatrix} \cdot & 0 & \cdot \\ 0 & \cdot & 0 \\ \cdot & 0 & \cdot \end{pmatrix}, \ \ \ \ \bs F_3 =\begin{pmatrix} \cdot & \cdot & 0 \\ \cdot & \cdot & 0 \\ 0 & 0 & \cdot \end{pmatrix}.
\end{gather*}
\caption{Classification of the 15 matrix textures with exactly two independent zeros.}
\label{tab:matrices}
\end{table}

We consider all charge assignments allowing for Majorana matrix textures with enough degrees of freedom to generate valid PMNS matrices. 
Since all permutations within the $a^3$, $ab^2$ and $abc$ classes are identical, we only need to consider one charge permutation for each class, while also setting $y^e=y^\nu$. The matrix $U_e$ is found by diagonalizing $y_ey^\dag_e$ while $W_\nu$ is found by first performing a singular value decomposition $m_\nu = A\Lambda B^\dag$, where $A$ and $B$ are both unitary and $\Lambda$ is real positive diagonal, so that $W_\nu = A(A^\dag B^*)^\frac{1}{2}$.  This algorithm is described in more detail in the Appendix~\ref{app:algorithm}.\\

For the $a^3$ class, the Majorana matrix can have any of the $39$ distinct invertible, symmetric matrix textures 
and give rise to a valid PMNS matrix. In the following we adopt the classification in Table~\ref{tab:matrices} for the discussed matrix textures~\cite{Fritzsch:2011qv}. For the $a^2b$ class with $T_L = T_\nu = (a,a,b)$, all invertible matrices result in a valid PMNS matrix except for the five textures that are subset textures of the Yukawa $y^\nu$ texture. The Yukawa texture can be chosen to be $\bs F_3$ as shown in Table~\ref{tab:matrices} and the excluded matrix textures are then given by 
\begin{align}\label{tab:aabclass}
\bs B_1\circ \bs F_3&= \begin{pmatrix} \cdot & \cdot & 0 \\ \cdot & 0 & 0 \\ 0 & 0 & \cdot \end{pmatrix},\quad\notag   
\bs E_3\circ \bs F_3=\begin{pmatrix} 0 & \cdot & 0 \\ \cdot & \cdot & 0 \\ 0 & 0 & \cdot \end{pmatrix}\\
\bs E_1\circ \bs F_3 &=\begin{pmatrix} 0 & \cdot & 0 \\ \cdot & 0 & 0 \\ 0 & 0 & \cdot \end{pmatrix},\quad   
\bs F_1\circ \bs F_3 =\begin{pmatrix} \cdot & 0 & 0 \\ 0 & \cdot & 0 \\ 0 & 0 & \cdot \end{pmatrix} \,,
\end{align}
where $\cdot$ is a general non-zero entry and $\circ$ is the entrywise (Hadamard) product. 

For the $abc$ class any texture with more than two independent zeros for the Majorana matrix fails to form a valid PMNS matrix, while all matrices with one or no zeros are allowed. The case when there are exactly two independent zeros requires a more subtle approach. Out of the 15 matrix textures classified in Table~\ref{tab:matrices},  all textures other than the $\bs F$ category lead to a sufficient number of degrees of freedom in the PMNS matrix. However, by directly linking the textures to the neutrino mass spectrum and the phases in a basis where the charged-lepton mass matrix is diagonal, it was found~\cite{Fritzsch:2011qv} that $m_\nu$ can only have textures of the $\bs A$, $\bs B$ and $\bs C$ categories. While for the $a^3$ and $a^2b$ classes, the previously allowed $m_\nu$ textures always have fewer than two independent zeros, this is not the case for two-zero Majorana matrix textures in the $abc$ class. Here, both the $\bs F$ and $\bs A$ categories are not allowed because they result in $m_\nu$ having a texture of the $\bs D$ category.

We are now in a position to determine all charge assignments resulting in Majorana matrix textures that allow for a valid PMNS matrix and, in conjunction with the conditions from anomaly cancellation,
to find all valid $U(1)_X$ groups formed with Majorana neutrinos.\footnote{Note that a similar classification of Abelian gauge groups leading to valid PMNS matrices and two-zero textures of $M^M$ has been presented in~\cite{Araki:2012ip}.} When one neutrino is massless, a valid PMNS matrix can only be found for the $a^3$ class, for which $M^M = 2C^k\langle \phi_k\rangle$ is a general symmetric matrix, whereas three massive neutrinos can also generate a valid PMNS matrix in the $ab^2$ and $abc$ classes. 

For the $a^3$ class, both $y_e$ and $y_\nu$ are completely general complex matrices and $U_e$ is a general unitary matrix. Since all neutrino charges are identical, $M_{ij}= 0$ for all $i,j=1,2,3$, and Majorana masses are not allowed unless at least one scalar with charge $q_{\phi} = -2a$ and $\langle \phi\rangle\neq 0$ is present. The Majorana matrix in this case becomes a general complex symmetric matrix and the anomaly condition relates the quark sector and lepton sector charges through $q_Q = -\frac{1}{3}a$, so this is another form of the $U(1)_{B-L}$ gauge group.\\
\begin{table}[t] \centering
\begin{tabular}{c c c c c}
\hline
$T_{\nu}$ & $b = -a$ & $b=0$ & $a=0$ & else  \\
$q_{\phi_k}$  & $-(a+b)$ & $-2b$ & $-2a$ & else \\
\hline
\hline
\addlinespace[1ex]
$M, C^k$ & $\begin{pmatrix} 0 & 0 & \cdot \\ 0 & 0 & \cdot \\ \cdot & \cdot & 0 \end{pmatrix}$ & $\begin{pmatrix} 0 & 0 & 0 \\ 0 & 0 & 0 \\ 0 & 0 & \cdot \end{pmatrix}$ & $\begin{pmatrix} \cdot & \cdot & 0 \\ \cdot & \cdot & 0 \\ 0 & 0 & 0 \end{pmatrix}$  & $\begin{pmatrix} 0 & 0 & 0 \\ 0 & 0 & 0 \\ 0 & 0 & 0 \end{pmatrix}$ \\
\hline
\end{tabular}
\caption{Assuming that $T_L = T_e = T_\nu = (a,a,b)$, here are the textures that can be formed for $M$ or $C^k$ given a specific choice of lepton charges or $q_{\phi_k}$, respectively.}
\label{aabMCtextures}
\end{table}
For the $ab^2$ class there are four allowed textures for $M$ and $C^k$, with the former depending on whether $a = 0$, $b=0$, or $a = -b$, while the latter depends on the charges $q_{\phi_k}$. The charge choices necessary to acquire the textures are given in Table \ref{aabMCtextures}. Since none of the $M$ matrices are rank-3, there is no valid $U(1)_X$ group for this class without at least one scalar. The most general case $T_L = (a,a,b)$ requires two scalars with charges $q_{\phi_1} = -(a+b)$ and $q_{\phi_2} = -2a$, with any more than two scalars being redundant. The only charge assignment that leaves the quark sector uncharged is $T_\nu = (a,a,-2a)$, $q_{\phi_1} = a$, $q_{\phi_2} = -2a$.\\

For the $abc$ class there are four ways to assign relative charges between $a$, $b$ and $c$ which lead to different $M$ textures. If all charges are different, $M=0$, but if $a=-b$, $a=-c$ or $b=-c$, then $M$ has a non-zero element in an off-diagonal position. If instead one of the charges is zero, then a diagonal element is non-zero. The previous two cases can also be combined to give a texture of a diagonal and an off-diagonal non-zero element. All these textures can also be constructed using differently charged scalars. An example of how to construct these three types of textures is given in Table \ref{abcMexample}.
\begin{table}[ht] \centering
\begin{tabular}{c c c c}
\hline
$T_{\nu}$ &  $b=-a$ & $c=0$ & $b=-a, c=0$ \\
$q_{\phi_k}$ & $-(a+b)$ & $-2c$ & $-(a+b)= -2c$ \\
\hline
\hline
\addlinespace[1ex]
$M,C^k$ & $\begin{pmatrix} 0 & \cdot & 0 \\ \cdot & 0 & 0 \\ 0 & 0 & 0 \end{pmatrix}$ & $\begin{pmatrix} 0 & 0 & 0 \\ 0 & 0 & 0 \\ 0 & 0 & \cdot \end{pmatrix}$ & $\begin{pmatrix} 0 & \cdot & 0 \\ \cdot & 0 & 0 \\ 0 & 0 & \cdot \end{pmatrix}$\\
\hline
\end{tabular}
\caption{This gives an example $M$ texture for each of the three 
different types of relative charge choices one can pick for $T_\nu = (a,b,c)$ that do not lead to a zero matrix.}
\label{abcMexample}
\end{table}
Adding the right number of scalars allows to create any texture.
One needs at least two scalars to create the first viable $abc$ class group $T_\nu = (a,-a, 0)$, four for the most general case of $T_\nu = (a,b,c)$ and three scalars for all other possible $T_\nu$ charge assignments. An explicit example of how the most general charge assignment with arbitrary $a$, $b$ and $c$ charges can be obtained is $q_{\phi_1} = -2a$, $q_{\phi_2} = -(a+b)$, $q_{\phi_3} = -2c$, and $q_{\phi_4} = -(b+c)$ giving a $\boldsymbol{B}_1$ texture for $C^k\langle \phi_k\rangle$, among many other possible charge choices. However, any texture of the $\bs F$ or $\bs A$ category must be avoided. The only charge choices that leave the quark sector neutral are $T_\nu = (a,-a,0)$ and $T_\nu = (a,b,-(a+b))$. Charged lepton family number differences, $U(1)_{L_i-L_j}$, are an example of this class~\cite{Foot:1990mn, Heeck:2011wj, Asai:2018ocx}.

%----------------------------------------------------
\begin{figure}[t]
    \centering
    \includegraphics[width=.5\textwidth]{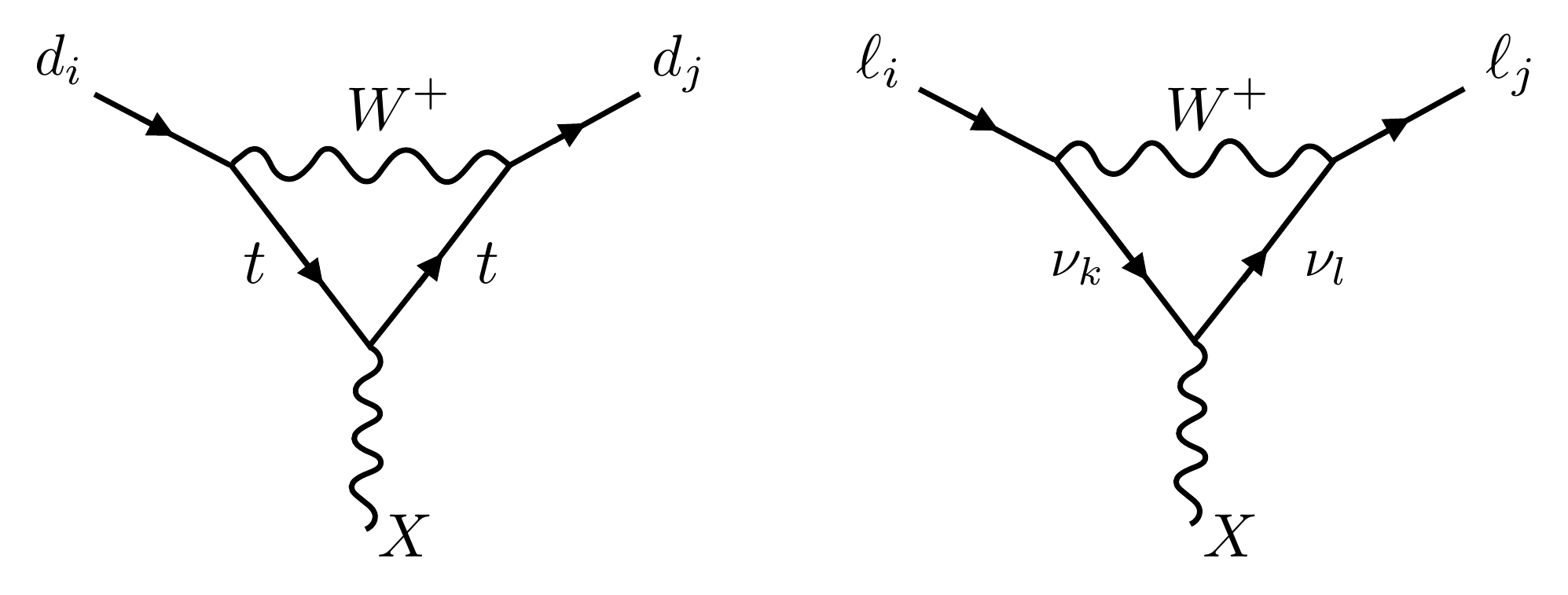}
    \caption{Loop diagrams contributing to flavour changing couplings of the $U(1)_X$ gauge boson. }
    \label{fig:feynman}
\end{figure}
%----------------------------------------------------

\section{Flavour changing Neutral Currents}\label{sec:fcnc}

An important difference between the couplings of the $X$ bosons of $U(1)_X$ in models with Dirac neutrinos and Majorana neutrinos is the flavour structure of $X$ boson couplings to neutrinos. 
In the presence of a Majorana mass matrix the rotation matrices $W_\nu$ and $U_\nu$ no longer commute with the charge matrix and we indicate the difference with \eqref{eq:com} by the index $M$, 
\begin{align}
[Q_\ell, W^M_\nu]= [Q_\nu, W^M_\nu]\neq 0\,,\\
[Q_\ell, U^M_\nu]= [Q_\nu, U^M_\nu]\neq 0\,,
\end{align}
unless the charge matrices $Q_\ell, Q_\nu$ are flavour universal.
Neutrino flavour changing interactions (in the neutrino mass basis) at tree-level are therefore a characteristic signature of minimal $U(1)_X$ extensions with Majorana neutrinos with the exception of the $U(1)_{B-L}$ group. In the case of $U(1)_{L_i-L_j}$ gauge groups the coupling to neutrinos is completely determined by the PMNS matrix $V_\text{PMNS}=U_\ell U_\nu^{M \dagger}=u_\ell U_\nu^{M \dagger} $, because $U_\ell =u_\ell \mathbbm{1}$ with a constant $u_{\ell}$. For $U_{B-3L_i}$ there is no unique dependence, but FCNCs are expected because $U_{\ell}$ is block-diagonal.  These neutrino flavour off-diagonal couplings are difficult to observe because neutrinos are produced coherently as flavour eigenstates and models with $X$ bosons that are non-diagonal \emph{in the flavour eigenbasis} are not minimal~\cite{Farzan:2015hkd, Babu:2017olk}. 

Even in the presence of neutrino FCNC at tree-level, the coupling matrix of the $X$ boson to charged leptons is diagonal. In principle, flavour changing couplings can be induced at loop-level, but are suppressed by a GIM mechanism. 
The corresponding loop diagram on the right in Fig.~\ref{fig:feynman} involves a sum that extends over different neutrino mass eigenstates in the loop as opposed to the case of the SM $Z$ boson, and the amplitude for an $i \to j$  transition is proportional to the factor
\begin{align}
[V_{\text{PMNS}}^\dagger \, (U_\nu  Q_\nu U_\nu^\dagger ) & V_{\text{PMNS}} ]_{ij }\notag \\
&= [U_l U_\nu^\dagger\, (U_\nu  Q_\nu U_\nu^\dagger )\, U_\nu U_l^\dagger]_{ij}\notag \\
&= (Q_\nu)_{ij}\,,
\end{align}
where in the last step  we used  \eqref{eq:com} and the fact that $Q_\nu = Q_\ell$. As a consequence, any charged lepton FCNCs are heavily suppressed by neutrino masses.

Flavour changing couplings to quarks are absent at tree-level as well. Even for a secluded $U(1)_X$ quark-flavour changing couplings arise from kinetic mixing with the SM hypercharge gauge boson. This kinetic mixing term is often divergent and the size of the mixing-induced flavour changing couplings are therefore dependent on the UV completion of these models~\cite{Bauer:2018onh}. In the absence of kinetic mixing, quark-flavour changing couplings are induced if the $X$ boson couples to the baryon current (for example in $U(1)_{B-L}$ and $U(1)_{B-3L_i}$). In light of the recent hints of lepton non-universality in $b\to s \ell^+\ell^-$ transitions, FCNCs in $U(1)_{B-3L_i}$ gauge groups are of particular interest. 
One of the Feynman diagrams inducing this coupling is shown on the left of Fig.~\ref{fig:feynman} for the case of external down-type quarks. Since the $U(1)_B$ charges are flavour universal, unitarity of the CKM matrix guarantees that any flavour changing coupling is proportional to the mass-squared differences of the internal quarks. Conservation of the $B$ current further requires flavour changing currents to be induced by higher-order operators which must vanish in the limit $M_X\to 0$ or depend on the external quark masses~\cite{Heeck:2014zfa}. The corresponding operators are
\begin{align}\label{eq:currents}
\mathcal{L}&= g^L_{ij} \frac{M_X^2}{M_W^2}\bar d_j \gamma_\mu P_L d_i X^\mu+ g^R_{ij}\frac{M_X^2}{M_W^2} \bar d_j \gamma_\mu P_R d_i X^\mu\notag\\
&+\frac{1}{2}g^\sigma_{ij}\bar d_j \sigma^{\mu\nu}\left(\frac{m_{d_j}}{M_W^2}P_L+\frac{m_{d_i}}{M_W^2}P_R\right)d_i X_{\mu\nu}\,,
\end{align}
with the couplings
\begin{align}
g^L_{ij}&=g_Xq_q \frac{\alpha}{8\pi s_w^2} V_{ti}V_{tj}^*f_1(x_t)\,,\\
g^R_{ij}&=0\,,\\
g^\sigma_{ij}&= g_X q_q\frac{\alpha}{8\pi s_w^2} V_{ti}V_{tj}^*f_2(x_t)\,,
\end{align}
where $q_q$ is the universal charge of the internal quarks, $x_t\equiv m_t^2/M_W^2$ and the loop functions $f_1(x_t)\approx 0.97$ and  $f_2(x_t)\approx -0.36$ can be found in the Appendix of~\cite{Inami:1980fz}. 
The new gauge boson $X$ can therefore be produced in flavour changing meson decays, where the vector and tensor currents in \eqref{eq:currents} induce decay widths proportional to the vector and tensor form factors, respectively
\begin{align}
\Gamma(B &\to K X)=\frac{1}{256\pi}\frac{M_BM_X^2}{M_W^4} \lambda_K^{3/2}\\
&\times\bigg[g_{32}^L M_B f_+(M_X^2) +g^\sigma_{32}m_bf_T(M_X^2) \Big(1+\frac{M_K^2}{M_B^2}\Big)^{-1}\bigg]^2\notag
\end{align}
where contributions proportional to $m_{d_i}$ are neglected and the form factors can be found in \cite{Bailey:2015dka, Straub:2015ica}. 
Flavour changing transitions with an $X$ boson are strongly suppressed. 
As a consequence, the contribution of a resonant, promptly decaying $X$ boson is constrained by searches in $B \to K$ transitions and 
with a mass of $250\, \text{MeV}<M_X<4700\, \text{MeV}$~\cite{Aaij:2016qsm}  
\begin{align}
\text{Br}(B^+ \to K^+ X\to K^+\mu^+ \mu^-)\lesssim 10^{-9} -10^{-10}\,,
\end{align}
In the following we present constraints and sensitivity ranges for current and future experiments for $U(1)_{B-3L_i}$ gauge bosons in the mass range relevant for flavour observables and compare the results of recent resonant searches in $B$ meson decays.

%----------------------------------------------------
\begin{figure*}[ht]
    \centering
    \includegraphics[width=.48\textwidth]{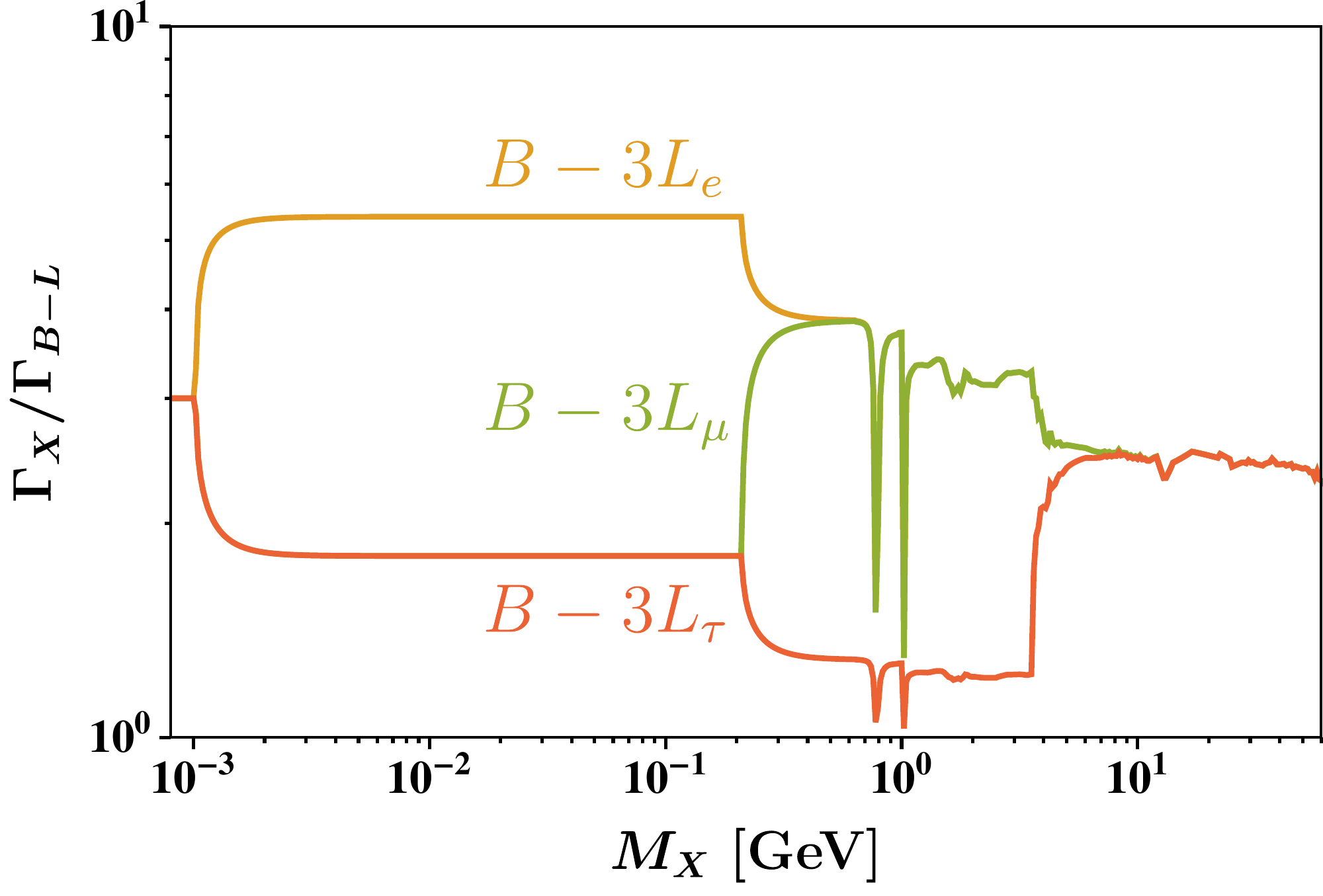}\hfill
    \includegraphics[width=.48\textwidth]{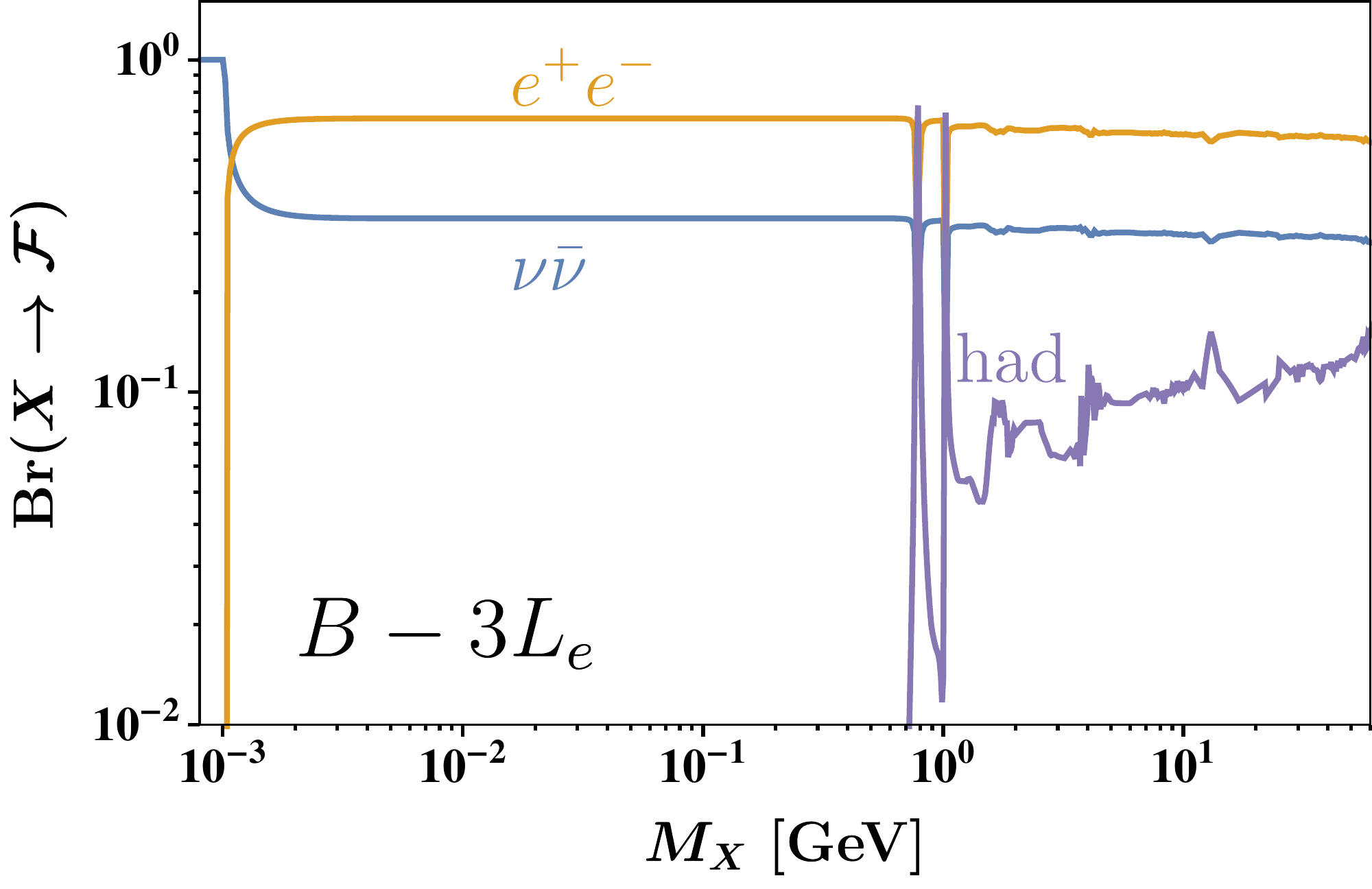}\\[4ex]
    \includegraphics[width=.48\textwidth]{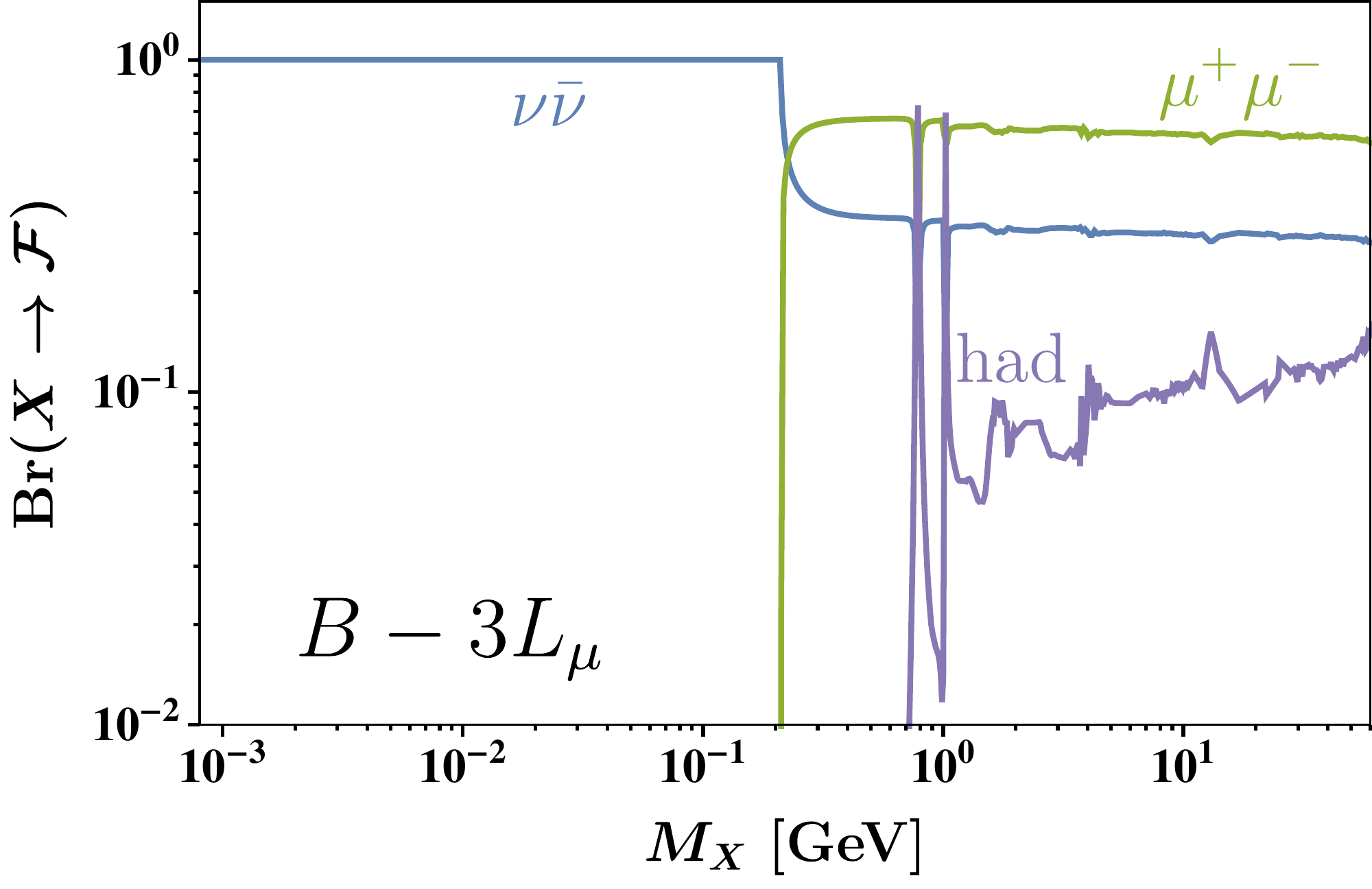}\hfill
    \includegraphics[width=.48\textwidth]{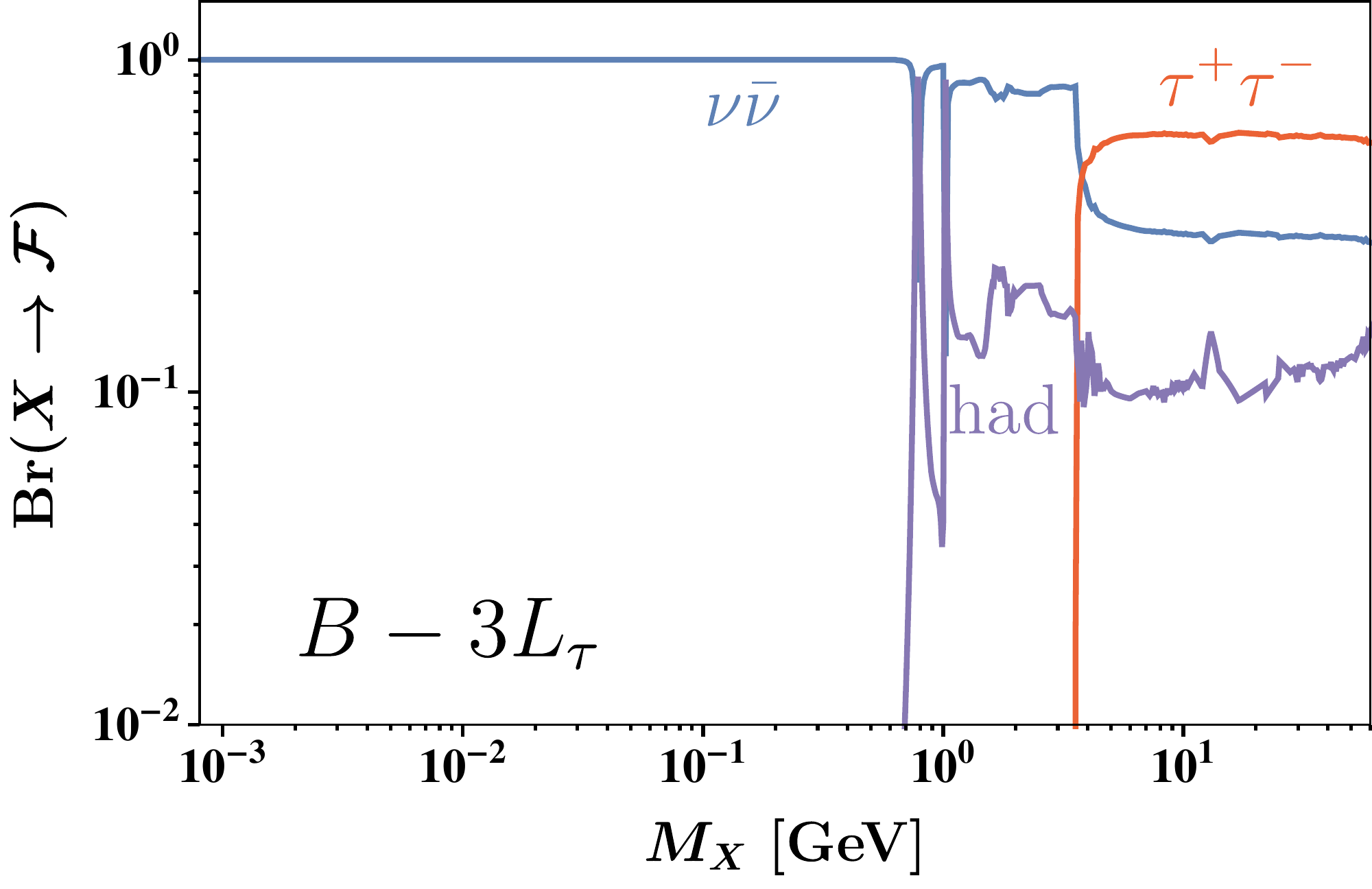}
    \caption{Total widths of the \Ubli{$i$} bosons normalised to the total width of the \Ubl boson (top left) without kinetic mixing. Branching ratios of the $U(1)_{B-3L_i}$ gauge bosons $X$ to different final states $\mathcal{F}$ in the case of $i=e$ (top right), $\mu$ (bottom left) and $\tau$ (bottom right) without kinetic mixing. }
    \label{fig:BRs}
\end{figure*}
%----------------------------------------------------

%%%%%%%%%%%%%%%%%%%%%%%%%%%%%%%%%%%%%%%%%%%%%%%%%%%%%%
\section{ Searching for $U(1)_{B-3L_i}$ Gauge Bosons }

The powerful suppression of quark and charged lepton flavour violating couplings are signatures of minimal, anomaly-free gauge groups with negligible kinetic mixing. This distinguishes minimal, anomaly-free gauge groups from extensions with anomalous currents~\cite{Dror:2018wfl, Dror:2020fbh}, a sizable kinetic mixing term, or the presence of additional scalars contributing to the fermion masses~\cite{Foot:1994vd}. 
The exception to this rule is the neutrino sector. Neutrino flavour changing interactions in the mass eigenbasis are absent for the \Ubl gauge boson, but expected at tree-level in any other minimal anomaly-free gauge boson corresponding to the gauge groups $U(1)_{L_i-L_j}$, $U(1)_{B-3L_i}$  and combinations thereof. It is challenging to measure these flavour changing currents. Very light vector bosons could be produced in active neutrino decays $\nu_i\to \nu_j X$~\cite{Barger:1981vd, Lessa:2007up}, but the corresponding parameter space is strongly constrained by measurements of the effective number of degrees of freedom $\Delta N_\text{eff}$, which are affected by the $X$ boson increasing the neutrino energy density through $X\to \bar \nu \nu$ decays~\cite{Escudero:2019gzq}.  In the following we will therefore focus on the mass range $M_X\gtrsim 10$ MeV for which constrains from $\Delta N_\text{eff}$ are substantially weaker. Searches for non-standard neutrino interactions (NSI) that are sensitive to neutrino couplings to matter (baryons or electrons) are excellent probes of non-universal and flavour changing interactions in this mass range. Importantly, in the flavour eigenbasis, couplings of all minimal anomaly-free gauge bosons are diagonal. Therefore, for the NSI Lagrangian
\begin{equation}\label{eq:eff_op}
    \mathcal{L}_\text{NSI} = -2\sqrt{2}\, G_F \sum_{f, \alpha, \beta} \varepsilon^{f}_{\alpha\beta} \ \left[\bar\nu_\alpha \gamma_\rho P_L \nu_\beta\right] \, \left[\bar f \gamma^\rho f \right]\,,
\end{equation}
where the sum extends over quarks and electrons $f=u,d,e$ and neutrino flavours $\alpha,\beta =1,2,3$,  the coupling structures for the different gauge groups are
\begin{align}
\epsilon^u_{\alpha\alpha}=\epsilon^d_{\alpha\alpha}=-3\epsilon^e_{\alpha\alpha}\quad &\text{for} \quad U(1)_{B-L},\\
\epsilon^e_{ii}=-\epsilon^e_{11}\quad &\text{for} \quad U(1)_{L_i-L_e},\,i=\mu,\tau\,,\\
\epsilon^u_{11}=\epsilon^d_{11}=-9\epsilon^e_{11}\quad &\text{for} \quad U(1)_{B-3L_e},\\
\epsilon^u_{ii}=\epsilon^d_{ii} \quad &\text{for} \quad U(1)_{B-3L_i},\,i=\mu,\tau \,.
\end{align}
While searches for NSI are sensitive to the different diagonal coupling structures of new minimal anomaly-free gauge bosons, they are not sensitive to the structure of the Majorana matrix reflected by flavour off-diagonal couplings to neutrinos in the mass eigenbasis. We discuss the strength of these constraints from various experiments for the three $U(1)_{B-3L_i}$ gauge bosons without kinetic mixing in detail and compare them with constraints from other experiments, complementing and updating previous analyses~\cite{Heeck:2018nzc,Kling:2020iar,Coloma:2020gfv}. Similar analyses have been performed for $U(1)_{B-L}$~\cite{Ilten:2018crw, Bauer:2018onh, Amrith:2018yfb, Kling:2020iar} and $U(1)_{L_i-L_j}$ gauge bosons~\cite{Ilten:2018crw,Bauer:2018onh,Amrith:2018yfb,Dror:2020fbh,Kling:2020iar,Coloma:2020gfv}. We further present the constraint from the resonance search in $B\to K\mu^+\mu^-$ decays for $U(1)_{B-3L_\mu}$ calculated in Section~\ref{sec:fcnc} and discuss the sensitivity required for future searches in meson decays to be competitive with existing constraints.\\

\begin{figure*}[ht]
    \centering
    \includegraphics[width=\textwidth]{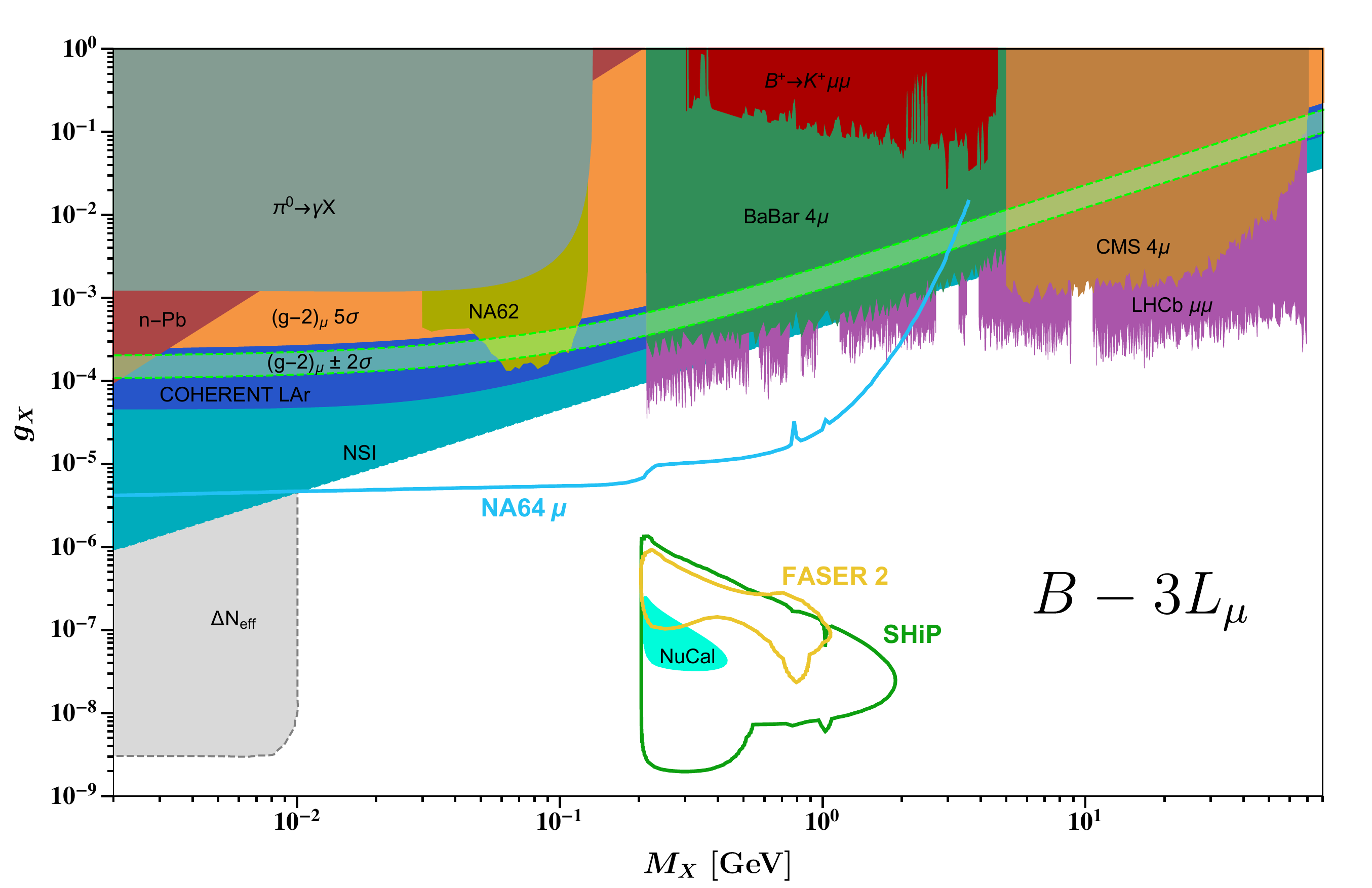}
    \caption{Limits on the parameter space of a  $U(1)_{B-3L_\mu}$ gauge boson without kinetic mixing. }
    \label{fig:lims_B3M}
\end{figure*}

In Fig.~\ref{fig:BRs} we show the decay widths and branching ratios for the \Ubli{$i$} gauge bosons and in Figs.~\ref{fig:lims_B3E} - \ref{fig:lims_B3T}, we show the two-dimensional parameter space defined by the gauge coupling $g_X$ and the mass $M_{X}$ of the \Ubli{$i$} gauge bosons in the absence of kinetic mixing. The shaded areas depict current bounds, while the coloured solid lines show future projected sensitivities.

\subsection{\Ubli{$e$}}

%----------------------------------------------------
\begin{figure*}[ht]
    \centering
    \includegraphics[width=\textwidth]{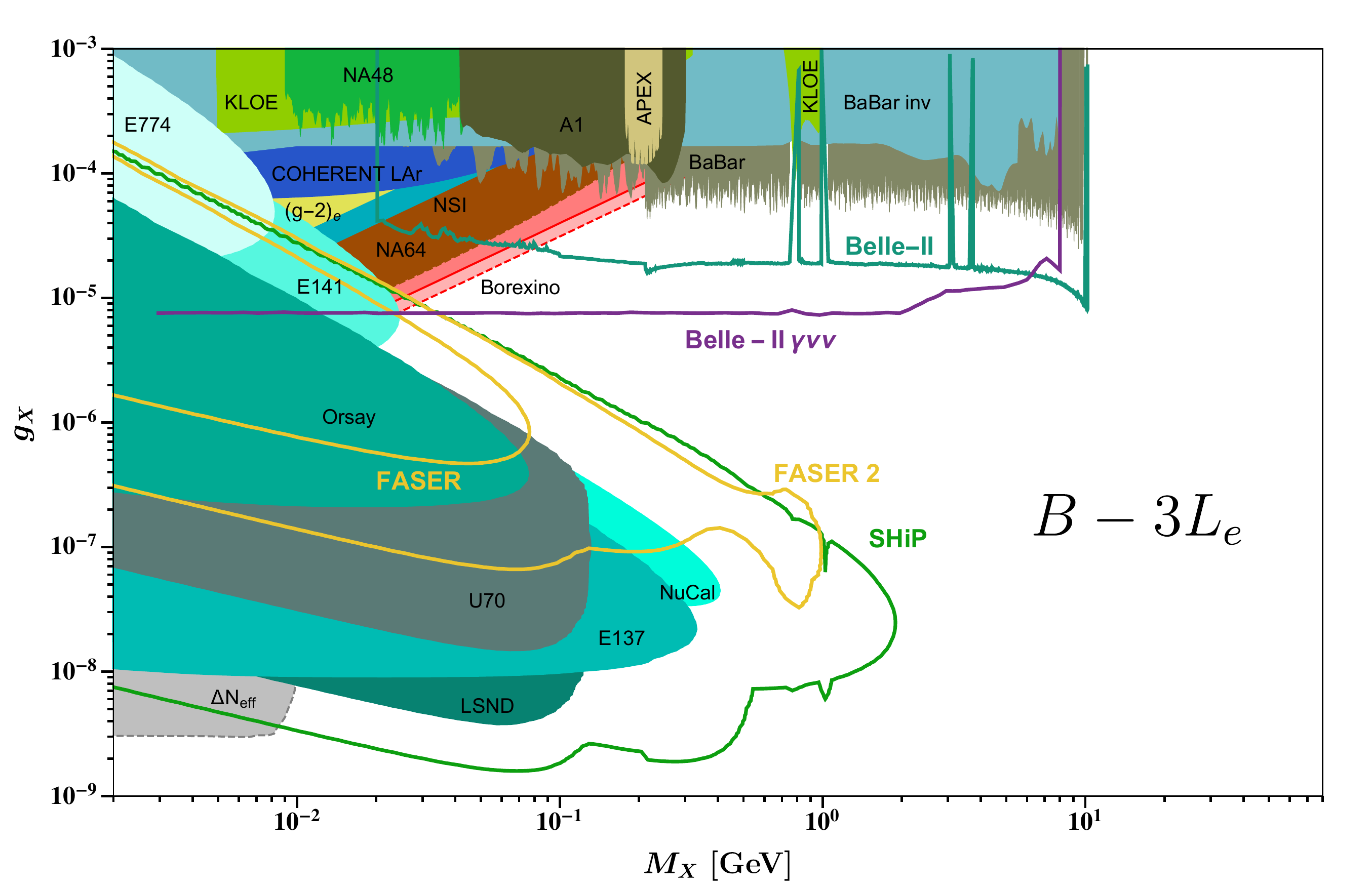}
    \caption{Limits on the parameter space of a  $U(1)_{B-3L_e}$ gauge boson without kinetic mixing. }
    \label{fig:lims_B3E}
\end{figure*}
%----------------------------------------------------

The phenomenology of a new \Ubli{$e$} gauge group summarised in the limits shown in Fig.~\ref{fig:lims_B3E} is overall very similar to  the case of $U(1)_{B-L}$, for which the corresponding exclusion limits can be found in~\cite{Ilten:2018crw, Bauer:2018onh}. This is readily understood as in both cases the new boson has gauge interactions with ordinary matter (i.e.~protons, neutrons and electrons), which for baryons are proportional to $B$ and for electrons to $L_e$. 

The fact that the charge of the electron is larger by a factor of 3 in the case of \Ubli{$e$} compared to \Ubl leads to a strengthening of the limits derived from searches of resonant hidden photon production in $e^+e^-$ colliders (like e.g.~A1~\cite{Merkel:2011ze,Merkel:2014avp}, APEX~\cite{Abrahamyan:2011gv}, BaBar~\cite{Lees:2014xha,Lees:2017lec}, KLOE~\cite{Archilli:2011zc, Babusci:2012cr, Babusci:2014sta}, NA64~\cite{NA64:2019imj})  by a similar factor. Furthermore, we  have computed constraints from searches at beam dumps and fixed target experiments (like e.g.~E137, E141, E774~\cite{ Bjorken:2009mm}, Orsay~\cite{Andreas:2012mt}, NuCal/U70~ \cite{Blumlein:2011mv,Blumlein:2013cua}, LSND~\cite{Essig:2010gu}) according to the analyses outlined in~ \cite{Bauer:2018onh}.
Due to the increased total width of the \Ubli{$e$} boson compared to the \Ubl case (cf.~top left panel of Fig.~\ref{fig:BRs}) the sensitivity of these experiments is shifted towards smaller couplings. This is because the larger width for a given coupling value $g_X$ corresponds to a shorter lifetime and hence the gauge boson decays earlier in the experimental apparatus.
Due to the absence of couplings of the new gauge boson $X$ to the second and third generation leptons, limits from searches in muonic final states like e.g.~the LHCb dimuon search are absent. Therefore, the \Ubli{$e$} gauge boson is rather unconstrained for masses above the Upsilon resonance, $M_X\gtrsim 10$ GeV.

Very low masses $M_X\lesssim 10$ MeV are strongly constrained through the allowed number of effective degrees of freedom $\Delta N_\text{eff}$ due to heating of the neutrino gas during neutrino decoupling in the early Universe~\cite{Escudero:2019gzq,Kamada:2015era}. The corresponding constraint shown by the light grey area was originally derived for the case of a $U(1)_{L_\mu-L_\tau}$ gauge boson coupling only to $\nu_\mu$ and $\nu_\tau$ neutrinos. We show it as a conservative estimate of the true bound for very low couplings $g\lesssim 10^{-5}$, which seems justified since the analysis of \cite{Escudero:2019gzq} including the effect of kinetic mixing (and therefore couplings to both baryons and electrons) shows that additional electron couplings rather weaken the bound in this regime.

Concerning neutrino scattering,  we have derived the constraint on a \Ubli{$e$} boson from the recently reported measurement of coherent elastic neutrino-nucleus scattering (\coherent) on liquid argon (LAr)~\cite{Akimov:2020pdx} at the COHERENT experiment. Therefore, we have performed the same $\chi^2$ analysis outlined as in \cite{Miranda:2020tif,Amaral:2020tga}.
The resulting bound shown in dark blue is not competitive.
The fact that this constraint is quite weak should not come as a surprise since the  \Ubli{$e$} boson is only coupling to the secondary electron-neutrinos of the COHERENT neutrino beam.
Similarly, the limit from neutrino oscillations in matter due to NSI~\cite{Coloma:2020gfv} shown in light blue is subdominant. 
A stronger bound arises from the missing energy search at NA64~\cite{NA64:2019imj} (shown by the brown area), which is sensitive to the invisible decays of the gauge boson $X$ into neutrinos. 

Most noticeably, the constraint from the Borexino experiment is the strongest current constraint for the parameter space not excluded by either accelerator and beam dump searches.
We have reanalysed the neutrino-electron scattering limits due to the Borexino Phase-I~\cite{Bellini:2011rx} and Phase-II~\cite{Agostini:2017ixy} determinations of the $^7$Be solar neutrino flux, following the analysis of~\cite{Amaral:2020tga}. This includes the full treatment of interference effects as well as the incorporation of previously ignored systematic uncertainties. The Phase-I  and Phase-II results are shown as the red dashed and solid lines, respectively.\footnote{For the derivation of these limits we have used solar neutrino fluxes as predicted by the Standard Solar Model~\cite{Vinyoles:2016djt} under the assumption that the Sun is a high metallicity star, as currently favoured by data.} We can view these two limits as the envelope of the true limit, which should be derived by combining the two individual results in  a single $\chi^2$ test taking into account any shared systematic uncertainties. 

In the future, novel fixed target experiments with high intensity beams like e.g.~SHiP~\cite{Anelli:2015pba,Alekhin:2015byh}  or the LHC forward detector FASER and the upgraded FASER 2~\cite{Feng:2017uoz,Ariga:2018uku} will have an improved sensitivity compared to current beam dump limits.\footnote{Here we want to thank Felix Kling for providing us with the meson spectra in the forward direction at LHC, which enabled a robust calculation of the FASER limits.} We calculated the corresponding projections and indicate the results by green and yellow solid contours, respectively,  in Fig.~\ref{fig:lims_B3E}. Furthermore, dielectron resonance searches as well as missing energy searches at the $e^+e^-$ collider Belle-II~\cite{Kou:2018nap} (illustrated by the cyan and purple contours, respectively) will have a significantly improved sensitivity over existing collider searches. 

Searches for resonances in $B\to K X\to Ke^+e^-$ decays would be sensitive to couplings $g_X\leq 10^{-4}$ for $M_X=1$ GeV if the experimental sensitivity  $\text{Br}(B\to K e^+e^-)\leq 10^{-15}$ could be achieved, at which such a search would probe unconstrained parameter space.

%----------------------------------------------------
%----------------------------------------------------

\subsection{\Ubli{$\mu$}}

In the case of \Ubli{$\mu$}, the two-dimensional parameter space shown in Fig.~\ref{fig:lims_B3M} looks quite different from the cases of \Ubl and \Ubli{$e$}. Focussing on beam dump and fixed target experiments first it is worthwhile noticing that below the dimuon threshold,  $M_X\lesssim 2 m_\mu$, beam dump and fixed target experiments are not sensitive, since the gauge boson cannot decay into any visible final states, as shown in the bottom left panel of Fig.~\ref{fig:BRs}. 
Therefore, only NuCal~\cite{Merkel:2014avp} has some sensitivity above the dimuon threshold, where the $X$ boson can decay into a pair of muons. This gives rise to the exclusion limit illustrated by the small cyan island.\\

Below the dimuon threshold, the gauge boson can be constrained from reactor neutron-nucleus scattering at keV energies~\cite{Barbieri:1975xy,Barger:2010aj}. The resulting limit from $n$-Pb scattering of $g_X\lesssim (M_{X}/206\ \text{MeV})^2$ is shown by the dark red area. Stronger constraints arise from a study of $\pi^0\to\gamma X$  decays  at \mbox{NOMAD}~\cite{Gninenko:1998pm,Altegoer:1998qta} excluding the dark grey area in Fig.~\ref{fig:lims_B3M}.
Searches for the decay $\pi^0\to\gamma\, (X\to\nu\bar\nu)$ at NA62~\cite{CortinaGil:2019nuo} provide a constraint for bosons with mass $M_{X}\sim 30 -130$ MeV displayed in dark yellow.
In this regime the $X$ boson is also constrained by the exclusion limit at the $5\,\sigma$ level from measurements of the muon anomalous magnetic moment $(g-2)_\mu$, which is shown by the orange area. The $2\,\sigma$ preferred region of the observed $(g-2)_\mu$ excess is illustrated by the light green band.
This is, however, ruled out by the bound from the COHERENT LAr run, which we derived in analogy to~\cite{Miranda:2020tif,Amaral:2020tga}. The reason this bound is so competitive in this model can be readily understood as the COHERENT neutrino beam consists mainly of a flux of prompt muon-neutrinos and secondary anti muon-neutrinos (apart from the secondary electron-neutrinos). 
 However, the strongest bound in this region of parameter space is provided by the limit on NSI-induced neutrino oscillations of~\cite{Coloma:2020gfv} illustrated by the dark cyan area. 
 At very low masses $M_X\lesssim 10$ MeV and couplings $g_X\lesssim 10^{-5}$ we again show the bound of~\cite{Escudero:2019gzq} in light grey  as an estimate for the constraint arising from $\Delta N_\text{eff}$.

Above the dimuon threshold, the $X$ boson can be constrained by resonance searches in muonic final states at various collider experiments. For example, peak searches in the dimuon invariant mass of a four-muon final state yield the green and dark orange exclusion limits from analyses by the BaBar~\cite{TheBABAR:2016rlg} and CMS~\cite{Sirunyan:2018nnz} collaborations, respectively. Bounds from a search for  prompt decays into a pair of muons at LHCb~\cite{Aaij:2019bvg} yield the strongest bound and are represented by the pink areas.
For comparison, we show the bound arising from the loop-induced flavour changing process $B^+\to K^+ \, (X\to \mu\mu)$ obtained by the resonance search at LHCb~\cite{Aaij:2016qsm}. The powerful suppression of 
quark FCNCs discussed in Sec.~\ref{sec:fcnc} renders this search non-competitive for the current experimental sensitivity. In order to probe currently unconstrained parameter space future searches would require a sensitivity of $\text{Br}(B\to K \mu^+\mu^-)\leq 10^{-15}$ for $M_X=1$ GeV.

Similar to the case of \Ubli{$e$}, we show that the sensitivity of searches at the future FASER  2 (yellow contour) and SHiP (green contour) experiments will be able to improve the limits from displaced searches in muonic and hadronic final states. Most noticeably, we find that missing energy searches with the planned upgraded run of NA64 with a dedicated muon beam will significantly push the limits towards smaller couplings $g_X$, shown by the light blue contour.

\subsection{\Ubli{$\tau$}}

%----------------------------------------------------
\begin{figure*}[ht]
    \centering
    \includegraphics[width=\textwidth]{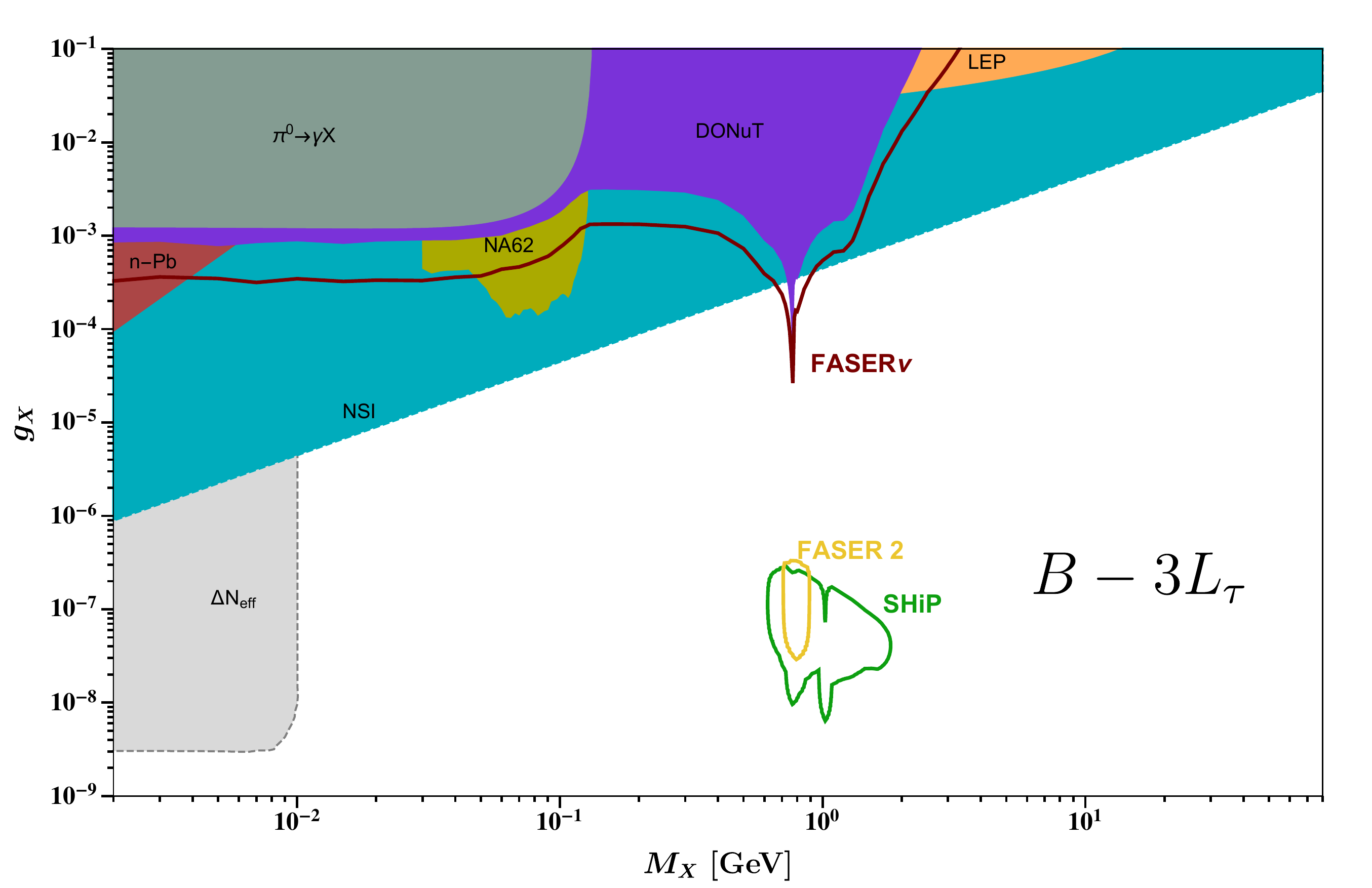}
    \caption{Limits on the parameter space of a  $U(1)_{B-3L_\tau}$ gauge boson without kinetic mixing.. }
    \label{fig:lims_B3T}
\end{figure*}
%----------------------------------------------------

The landscape of constraints on \Ubli{$\tau$} shown in Fig.~\ref{fig:lims_B3T} alters significantly from all previously discussed cases. To begin with, current and past fixed target and beam dump experiments relying on visible decay products (i.e.~electrons, muons and hadrons)  are not sensitive to this scenario. This is due to the absence of first and second generation lepton couplings in this model, as well as the fact that the energies have not been high enough to produce hadronic final states at past experiments.

However, in \cite{Kling:2020iar} limits on \Ubli{$\tau$} have been derived from a number of experiments. In particular, scattering of tau neutrinos produced in the decay of $X$ bosons with the DONuT~\cite{Kodama:2007aa} detector can be used to constrain this model. These limits obtained in~\cite{Kling:2020iar} are shown by the purple area. The previously mentioned constraints from neutron scattering is shown in dark red, and the constraints from semi-visible pion decays at NOMAD and NA62  are shown in dark grey and dark yellow, respectively. 
Finally, $Z$ lineshape measurements at LEP constrain any new physics contribution to the partial width of the $Z$ into taus~\cite{Ma:1998dp} and result in the bound on the $X$ boson shown by the orange area.

The dominant constraint on this model, however, is due to NSI-induced neutrino oscillations~\cite{Coloma:2020gfv}, which is shown by the cyan area. This bound is more stringent than the previously mentioned limits by up to two orders of magnitude.
At very low masses $M_X\lesssim 10$ MeV we again show the bound of~\cite{Escudero:2019gzq} in light grey  as an estimate for the constraint arising from $\Delta N_\text{eff}$.

Furhter, we show that in the future FASER 2 (yellow contour) and SHiP (green contour) will have some sensitivity to much smaller couplings as constrained by NSIs for GeV-scale gauge boson masses  due to searches of long-lived decays into hadronic final states. Furthermore, the recently proposed FASER$\nu$ experiment will be sensitive to scattering of tau-neutrinos produced at the LHC~\cite{Abreu:2019yak, Abreu:2020ddv}. This will enable FASER$\nu$  (red contour) to improve the limits over those previously set by NSI in a small mass window around the $\omega$-resonance. Similar to the resonance search for muons in $B$ meson decays a future resonance search with tau final states $B\to K X\to K\tau^+\tau^-$ could probe unconstrained parameter space if an experimental sensitivity of  $\text{Br}(B\to K e^+e^-)\leq 7.4\times10^{-14}$ could be achieved. This search would then be senitive $g_X\leq 10^{-3}$ for $M_X=4$ GeV.

%%%%%%%%%%%%%%%%%%%%%%%%%%%%%%%%%%%%%%%%%%%%%%%%%%%%%%
\section{Conclusions} 

In this work we have systematically studied the possible coupling structures of the associated boson of a new $U(1)_X$ gauge group to SM fields including three right-handed neutrinos. Constraints from gauge anomaly cancellation, fermion masses and mixing angles allow only for a limited number of $U(1)_X$ groups. We categorise these extensions and determine the allowed textures of Yukawa and Majorana mass matrices in all cases. If neutrinos are Dirac particles, the only possible choice satisfying all conditions is the $U(1)_{B-L}$ group, whereas all other $U(1)_X$ extensions require Majorana masses. We show that quark and charged lepton flavour transitions are strongly suppressed independent of the gauge group, but flavour changing couplings of the $X$ gauge boson to active neutrinos in the mass basis provide a signature capable to distinguish between the $U(1)_{B-L}$ group and \emph{any} other minimal, anomaly-free $U(1)_X$ extension of the SM. The structure of these couplings is determined by the underlying gauge group and the texture of the corresponding Majorana mass matrix. Existing experimental searches are not sensitive to flavour changing neutrino transitions in the mass basis, but provide constraints on flavour non-universal couplings through non-standard interactions if couplings to baryons or electrons are present.  We calculate the loop-induced quark flavour changing couplings of new gauge bosons coupling to the baryon current and present the constraint on the $U(1)_{B-3{L_\mu}}$ gauge boson from recent resonance searches for $B^+\to K^+ \mu^+\mu^-$ at LHCb. We perform an extended and updated analysis and contrast current constraints and the sensitivity reach of proposed experiments for the $U(1)_{B-3{L_e}}, U(1)_{B-3{L_\mu}}$ and $U(1)_{B-3{L_\tau}}$ gauge bosons for neutrino observables and quark flavour transitions with beam dump, collider and precision experiments.

%%%%%%%%%%%%%%%%%%%%%%%%%%%%%%%%%%%%%%%%%%%%%%%%%%%%%%
\section*{Acknowledgements} 

MB thanks Uli Haisch and Aleksej Rusov for helpful discussions on flavour changing couplings of hidden photons. PF wants to thank Pilar Coloma and Felix Kling for helpful discussions and for providing us with valuable data. Furthermore, PF wants to express his gratitude to Mirjam G\"{o}lz, Richard K\"{o}nigsdorfer, Katharina Lutz and Aaron von Siebenthal for their hospitality during these difficult times. PF acknowledges funding by the UK Science and Technology Facilities Council (STFC) under grant ST/P001246/1.

\appendix

\section{Diagonalisation}\label{app:algorithm}

The validity of the PMNS matrix for each charge assignment was checked by considering random matrices with the required texture and explicitly calculating its degrees of freedom. We used the standard parametrisation in terms of Euler angles and phases given by
\begin{equation}
V = \begin{pmatrix} e^{i\delta_e} & 0 & 0 \\ 0 & e^{i\delta_\mu} & 0 \\ 0 & 0 & e^{i\delta_\tau} \end{pmatrix} \tilde V \begin{pmatrix} e^{-i\phi_1/2} & 0 & 0 \\ 0 & e^{-i\phi_2/2} & 0 \\ 0 & 0 & $1$ \end{pmatrix},
\end{equation}
where
\vspace{-1ex}
\begin{align}
& \tilde V = \\
& \begin{pmatrix}
c_{12}c_{13} & s_{12}c_{13} & s_{13}e^{-i\delta} \\
-c_{23}s_{12} - s_{23}s_{13}c_{12}e^{i\delta} & c_{23}c_{12}-s_{23}s_{13}s_{12}e^{i\delta} & s_{23}c_{13} \\
s_{23}s_{12} - c_{23}s_{13}c_{12}e^{i\delta} & -s_{23}c_{12}-c_{23}s_{13}s_{12}e^{i\delta} & c_{23}c_{13}
\end{pmatrix},\notag
\end{align}
and $c_{ij} = \cos \theta_{ij}$ and $s_{ij} = \sin \theta_{ij}$. The Euler angles are calculated from the matrix via
\begin{align}
&\theta_{13} = \sin^{-1}(|V_{13}|), 
\end{align}
\begin{align}
& \theta_{12} = 
\left\{ \begin{array}{ll} 
\tan^{-1}\frac{|V_{12}|}{|V_{11}|} & \mbox{if} \ V_{11} \neq 0 \\
\frac{\pi}{2} & \mbox{else},
\end{array}
\right. 
\end{align}
\begin{align}
& \theta_{23} = 
\left\{ \begin{array}{ll} 
\tan^{-1}\frac{|V_{23}|}{|V_{33}|} & \mbox{if} \ V_{33} \neq 0 \\
\frac{\pi}{2} & \mbox{else},
\end{array}
\right.
\end{align}

while the six phases are found using \cite{Antusch:2003kp}
\begin{align}
& \delta_\mu = \text{arg}(V_{23}), \\
& \delta_\tau = \text{arg}(V_{33}), \\
& \delta = -\text{arg}\bigg( \frac{V^*_{11}V_{13}V_{31}V^*_{33}}{c_{12}c^2_{13}c_{23}s_{13}s_{12}s_{23}} + \frac{c_{12}c_{23}s_{13}}{s_{12}s_{23}}\bigg), \\
& \delta_e = \text{arg}(e^{i\delta}V_{13}), \\
& \phi_1 = 2 \text{arg}(e^{i\delta_e}V^*_{11}), \\
& \phi_2 = 2 \text{arg}(e^{i\delta_e}V^*_{12}).
\end{align}

The angles and phases were considered to be valid as long as they were not multiples of $\pi$ and $\frac{\pi}{2}$, respectively. For the PMNS matrix to be considered compatible with the SM measurements, we required all three angles and at least one physical phase to take non-trivial values. We also do not require the presence of two additional Majorana phases since they have not been experimentally confirmed.

%%%%%%%%%%%%%%%%%%%%%%%%%%%%%%%%%%%%%%%%%%%%%%%%%%%%%%

\bibliography{literature}

\end{document}